\documentstyle[aps,preprint,12pt]{revtex} 

\begin{document}
\draft

\title{Level spacing statistics of disordered finite superlattices spectra and  
motional narrowing as a random matrix theory effect }

\author{R. R. Rey-Gonz\'alez}
\address{Departamento de F\'{\i}sica, Universidad Nacional de Colombia\\
Santaf\'e de Bogot\'a, Colombia\\}

\author{and\\ P. A. Schulz}
\address{Instituto de F\'{\i}sica Gleb Wataghin, UNICAMP, Cx.P. 6165, \\
13083-970 Campinas SP Brazil}

\maketitle

\date{today}

\begin{abstract}
In the present work the problem of coupled 
disordered quantum wells is addressed in 
a random matrix theory framework. The quantum wells are short repulsive 
binary alloys embeded by ordered barriers and show well defined 
quantized levels as a consequence of spatial confinement. 
 Finite disordered superlattices may show both diffusive-like and 
localized minibands.
Three different level repulsion suppression mechanisms
 are discussed by analysing the evolution of 
 nearest-level-spacing distribution function 
 within each superlattice miniband.
  The present numerical results show a motional narrowing 
 effect, which is in fact a consequence of the random matrix theory.

\noindent PACS number(s) 71.23.An,73.20.Jc,74.40.+k,73.20.Dx

\end{abstract}

\newpage 

\section{Introduction}

In a recent work we discuss a model disordered quantum well made of a
short repulsive binary alloy embeded by ordered barriers\cite{rey1}.  
Since repulsive
binary alloys show a correlation in the disorder, an effective
delocalization window could be established in the energy range below the
barrier top.  Therefore, the intensively studied problem of delocalization
in one-dimensional chains with correlated disorder 
\cite{flores,dunlap,wu,sen,datta,chen,evangelou,sanchez,heinrichs,soukoulis}
 could be addressed from
the point of view of quantization effects due to spatial confinement in
such chains.   A further motivation concerns the verifiability of quantum
confinement effects in amorphous semiconductor heterostructures, a largely
unanswered question\cite{chun,hattori}.  In this context repulsive binary alloy quantum wells
lead to a description of these systems including inherently a key feature
of many disordered bulk materials:  the presence of well defined energy
ranges showing either localized or effectively delocalized states\cite{rey2}.  

The existence of well defined spatial confinement quantization in these 
model disordered quantum wells poses a new question concerning the coupling 
of quantum wells in finite superlattices. Although well defined, the quantum 
well states show an inhomogeneous broadening and therefore it is not clear how
the level repulsion due to quantum wells coupling behave as a function of the 
localization length. Once the coupling of two disordered quantum wells by 
means of tunneling through a barrier could be described, the evolution of level 
repulsion in finite superlattices could show still 
qualitative different aspects. 

In the present paper we show that the coupling of two quantum wells can be well 
described by the usual tunneling mechanism in a picture delivered by appropriate 
level spacing statistics for an ensemble of double quantum wells. Furthermore,
the level spacing statistics evolve to Poisson or Wigner surmise distributions 
\cite{izrailev90}
in finite superlattices, depending on the miniband index. 
For sake of clearness, in the present work we will name as a miniband the level 
clustering around the average energy of the levels of isolated quantum wells.
Hence, level spacing statistics will be applied for levels within a given 
miniband. 
Therefore, for a 
given set of parameters, while one level cluster may show signatures of true 
minibands \cite{bastard}, 
others present from an effectively extended behaviour down to strong 
localization in 
the same superlattice. The minibands of effectively extended states have 
level spacing properties of a diffusive-like regime. 
We also identify in the present system a motional 
narrowing effect in difusive-like minibands, which is actually a consequence of
random matrix theory (RMT)\cite{weiden}.

In what follows we first describe briefly the model Hamiltonian for the 
disordered quantum wells and finite superlattices, as well as the level spacing 
statistics approach. Afterwards results for finite disorder superlattices 
spectra will be shown and discussed based on the nearest neighbour level spacing 
statistics properties. Finally, the numerical results for the motional narrowing 
effect will be presented and analysed as a consequence of RMT.

\section{Disordered finite superlattice model Hamiltonian}

The present model Hamiltonian consists of a one-dimensional chain of {\it
s}-like orbitals, treated in the tight-binding approximation, with
nearest-neighbor interactions only,

\begin{equation} H=\sum_{n}({\varepsilon_{n}}|n><n| + V_{n,n+1}|n><n+1| +
V_{n+1,n}|n+1><n|), \end{equation}

A finite chain segment will emulate a single well, or a finite superlattice,
 sandwiched by  infinite barriers (isolated structure). The actual chain is 
 constructed in a way that short disordered segments representing the quantum 
 wells are separated by sets of few sites for the barriers. The ends of the 
 total chain segment will emulate wider barriers,
  of same high as the internal ones, in order to minimize surface 
 effects at the quantum wells at the
 ends of the finite superlattices.

 The well material is the above mentioned repulsive binary
alloy\cite{dunlap,wu}, where the bond between one of the atomic species is
inhibited, introducing short range order: in a chain of {\it A} and {\it
B} sites, only {\it A-A} and {\it A-B} nearest neighbors bonds are allowed.
The introduction of this short range order leads to delocalization of
states in the disordered chain\cite{soukoulis,rey2}. The well layer is then
characterized by the correlation in disorder and the concentration of {\it B}-like sites, which is
related to the probability $P_B$ of a {\it B}-like site to be the next one
in generating a particular chain configuration.  All results shown here are
for $P_B = 0.5$, corresponding to an effective concentration of $B-$like
sites of $\rho_B \approx 0.3$.  Disorder is straighfowardly introduced by
randomly assigning {\it A} and {\it B} sites, according to the constraints
on concentration and bonding mentioned above.
 Since we are simulating
disordered systems, averages over hundreds of configurations for the same 
parameters are undertaken.

The atomic site energies used throughtout this work are $\epsilon_A = 0.3$
eV, $\epsilon_B = -0.3$ eV and the hopping parameter are $V_{AA} =-0.8$ eV
between {\it A}-like sites and $V_{AB} =-0.5$ eV between {\it A}-like and
{\it B}-like sites for the well layer\cite{rey2}. For the barriers these
parameters are $\epsilon_{br} = 0.8$ eV and $V_{br} =-0.4$ eV. The results 
shown here are all considering wells of $L_{w} = 40$ sites and  barriers widths 
of $2 \le L_{b} \le 15$ sites.  
The external barriers are characterized by the same barrier parameters given 
above and are $L_{ext} = 11$ sites wide. At interfaces we consider
 geometric averages of
the hopping parameters. The number of quantum wells in a finite superlattice is 
varied in the range $2 \le N_w \le 19$, throughout the present work.  

\section{Level spacing statistics}

A bona fide quantum well state in a single disordered quantum well 
still shows an inhomogeneous broadening due to the underlying disorder,
 related 
to a finite localization length, even if this length is many times longer
 than the well 
width, a condition necessary for the spacial quantization itself. On the other 
hand, 
considering two coupled quantum wells, it is important to compare
the energy scale of this inhomogeneous 
broadening to the tunneling spliting of the levels. A
 level spliting must be determined as an average over 
 several double well like chains with different disorder configurations.
 However, if the broadening is of 
the order of (or larger than) this spliting, then the coupling of the 
quantum wells can not be resolved numerically in an average of the energy 
spectra. We will see that this is actually the most common situation we found 
by numerical inspection of the problem. Therefore we must consider a nearest 
neighbour level spacing statistics of the spectra. 

Having in mind the 
double-well system, we refer to the coupling between the quantum wells as a two
level problem \cite{bohr} and the splitting S is given by 

\begin{equation}
S^2=(H_{11}-H_{22})^2 + (2 \cdot H_{12})^2
\end{equation}

where $H_{11}$ and $H_{22}$ are the energies of equivalent quantized levels,
 one in each 
well, while $H_{12}$ is the coupling between both wells. 
If $H_{11}-H_{22}$ and $2H_{12}$ are 
independent Gaussian random variables, the distribution of the level spacing 
(tunneling
spliting) is given by a Wigner surmise:

\begin{equation}
P(S) = {{\pi}\over 2D^2}S~\exp- \left\{ {{\pi}\over 4} {S^2\over D^2}\right\}
\end{equation}

where $D$ is the average spacing. If $H_{12} = 0$, i.e, 
there is no coupling between 
the states of the two quantum wells, the distribution of the level spacing 
reduces to the Poisson distribution:

\begin{equation}
P(S)={1\over D}~\exp - \left\{ {S\over D} \right\}
\end{equation}

The average level spacing, $D$, is related to the variance, $\sigma^2$, of 
the Gaussian random variables by $D=\sigma\sqrt{{\pi\over 2}}$ \cite{bohr}.
These are two interesting limits given by RMT which establish 
a useful investigation tool for the coupling between two (or more) disordered 
quantum wells, relative to the localization length in these systems 
\cite{shapiro}. 
  It should be noticed that, while
$H_{11}-H_{22}$ is a random variable, since there is a inhomogeneous broadening
 of the levels; $2H_{12}$ is not entirely random for a few quantum wells system.
  In what follows we will show that numerical results confirm 
 this prediction, but the coupling between wells will be independently 
 randomized in multiple coupled quantum wells and the Wigner surmise will 
 also become an important limit of the general problem.

 The level spacing statistics for levels within a 
given miniband is a well defined problem, since the clustering of eigenvalues 
in minibands delivers sets of equal 
number of levels of comparable localization lengths and these sets are well 
separated in energy from each other. In other words, a finite superlattice 
overcomes the problem of arbitrarily defining energy intervals where to evaluate 
the level spacing distributions, like in long one-dimensional chains with 
correlated disorder\cite{russ} showing an almost continuous
 spectra of states with 
localization lengths that are strongly energy dependent\cite{izrailev96}. 

\section{Energy spectra and level spacing distributions}

Previous results on single disordered quantum wells \cite{rey1} show
 that quantum confinement 
effects are already resolved 
if some average level spacings are greater than the level broadening.
However, the formation of superlattice minibands with the successive coupling 
of disordered quantum wells is related to a different energy scale.
In order to build up a miniband of Bloch states, the inhomogeneous level 
broadening has to be much less than the level repulsion due to the tunneling 
coupling. For a given tight-binding parameters set this quantum well level 
spliting can be tuned by changing the barrier width. After several numerical 
tests we chose the parameters listed in section II. For the most 
favorable situation 
found for barrier widths down to $L_b = 2$ sites, the average level repulsion 
turned out to be of the order of the inhomogeneous broadening of the levels, 
with the exception of a peculiar quantum well level, as will be discussed below.

In Fig.1 we show the energy spectra 
as a function of disorder configuration, 
for a single quantum well, a double quantum 
well and a finite superlattice of five quantum wells, for $L_b = 2$. The 
spectra are shown in the range around the energy of maximum localization length 
of an infinite repulsive binary alloy with the given parameters, $\lambda_{max}
\approx -0.7 eV$ \cite{rey2}. We can see an internal structure for the level 
$n=12$, that partially survives in the double quantum well and in 
the finite superlattice, with clear signatures of level spliting due to 
quantum well coupling. For a finite superlattice this intra-miniband structure 
is better observed in the level-spacing distibution, as will be seen below.
 This quantum well level is closely tuned to the 
$\lambda_{max}$ energy for a $L_w = 40$ sites well. Hence, the level broadening 
is the lowest one, leading to the resolution of the internal structure of the 
average level spectrum. This structure is related to the fact that the well is 
a repulsive binary alloy and the effective well width varies according to the
interface sites. Three different configurations at the interface are possible:
{\it (a)} A-like site at both interfaces, 
{\it (b)} one A-like site at one interface and a B-like 
at the other; {\it (c)} and B-like sites at both interfaces. The localization 
increases from {\it (a)} to {\it (c)}, proportional to the 
detuning of the ${\lambda_{max}}$ condition and the consequent increase in the 
inhomogeneous broadening. The level repulsion for these $n = 12$ states is 
partially resolved on average for a double-well. Nevertheless, increasing 
further the number of wells, the average energy separation of nearest neighbour 
levels within a miniband decreases and the coupling among quantum wells is not
resolved anymore for a finite superlattice, as can already be seen for finite 
superlattice of five quantum wells. The miniband imediately below, related to 
the level $n =11 $, has an inhomogeneous broadening already wider than the 
internal average structure due to the interface configuration. On the other 
hand, these states are still bona fide quantum wells states and should repeal
each other by quantum well coupling, but such level spliting is not resolved 
in the average spectrum even in the case of a double-well with very thin 
barriers, Fig.1.

The level splitting for these states may actually be 
quantified in histograms of individual levels of the $n =  11$ doublet, Fig. 2.
In Fig. 2(a) the histograms generated by the lower and the higher state taken
separately in 1000 disorder configurations are depicted. Each level shows 
a nearly Gaussian inhomogeneous broadening with a half width of the order of 
the average level spliting given by the energy separation of curve peaks, 
$\delta$. 
The 
average density of states in this energy range, Fig. 2(b), on the other hand,
 is 
structureless, as mentioned above. Since 
disordered 
quantum wells states show a coupling, but with a level spliting of the order
 or less than the individual level broadening, the density of states is not a
 useful quantity to be analised and one should look at the nearest neighbour 
 level spacing statistics.
 
  Nearest neighbour level spacing probabilities for double disordered 
 quantum wells are shown in Fig.3. We consider the $n=12$, Fig. 3(a); and 
 $n=11$, Fig. 3(b), doublets. The numerical histograms, obtained taking 
 into account 1000 disorder configurations, are compared to analytical Poisson
 and Wigner distributions for the numerically obtained average level spacing,
 $D$. Here we also consider
  different barrier thicknesses, $L_b = 2$ and $L_b = 5$ sites wide, as a 
  mechanism that modify the level repulsion. First, for thin barriers
   (strong coupling),
   top of Fig. 3(a) and 3(b), we see that the level spacing probabilities 
   present a threshold, $\Delta$, for both doublets. This threshold, i.e., the 
   minimum value for level spacing, is a direct measure for the tunneling 
   coupling between resonant states. This can be confirmed by inspecting that 
   indeed $ln\Delta \propto -L_b$. The spacing probability for the $n=12$ doublet 
   show a rich structure above the threshold, which is due to the resolved 
   interface related structure, already seen in the energy spectra as a function 
   of disorder configuration, Fig.2. For the $n = 11$ states the spacing 
   probability distribution is a smoother structureless curve, which is an 
   evidence for an level broadening wider than the level repulsion.
   
   This picture changes completely for thicker barriers. 
   By increasing the barrier thickness, the minimum level spliting diminishes 
   exponentially, becoming neglegible compared to the average level spacing. In 
   this limit, already seen for $L_b = 5$, the nearest neighbour level spacing 
   approaches a Poisson distribution, as can be seen 
   in the bottom panel of Fig. 3(b), which is 
   expected for uncorrelated levels, localized in either one of the wells. In 
   the bottom part of Fig. 3(a), for the $n=12$ (more delocalized) states, there 
   are still structures in the level spacing distribution and no definitive 
   answer concerning the correlation of levels can be delivered from comparisons 
   with either Wigner surmise or Poisson distributions.   
   Interesting to notice is the fact that no Wigner surmise like distribution 
   is obtained in the thin barrier limit. Although the almost degenerate levels 
   show a Gaussian random distribution, the coupling between them is not random. 
   The finite threshold shown in
    the top of Fig. 3(b) actually suggests that the coupling is 
   nearly constant in a first order approximation. 
   
   Inter-well coupling, however, 
   may be randomized if we increase the system by adding successive quantum wells,
   generating a finite superlattice. With such a procedure, the level repulsion
   diminishes linearly and not exponentially and a Wigner like distribution 
   could be expected.
   This trend can be already observed for a finite superlattice of five quantum 
   wells, keeping the barrier width $L_b = 2$, Fig. 4. 
   In this figure the level spacing for other minibands of 
   more localized states are also depicted: (a) for the 
   $n=12$ miniband; while (b), (c), and (d) are for the 
   $n=11$, $n=10$, and $n=9$ minibands, respectively. Now we see that the 
   miniband related to $n = 12$, although showing some structure, presents a 
   clear Wigner surmise envelope. 
   For low spacing, however, a threshold is still
    seen. This threshold is already absent for the other minibands in this five 
    quantum wells finite superlattice. 
    For the $n=11$ states related miniband, the numerical 
   simulation of level spacings
    shows a perfect agreement with the Wigner surmise distribution. This 
    agreement is still reasonable for the $n=10$ miniband, Fig. 4(c), while for
    lower miniband indexes, like in Fig. 4(d), a crossover from Wigner surmise 
    to Poisson distribution is clearly identified. 
   
   The fact that in a finite superlattice of disordered quantum wells, nearest 
   neighbour levels associated to the same minibands obey a Wigner surmise 
   indicates that these levels are still correlated and extended along the 
   entire system, Fig. 4(b). Besides, we may determine a lower 
   bond for the localization 
   length of these states: as far as the level spacing in a given miniband 
   follows the Wigner surmise, the localization length exceeds the system 
   length, $\lambda > L$ \cite{izrailev90}. 
   Based on this evidence, further increasing of the
   system length, by adding more quantum wells, would lead to a transition to 
   localization, i.e., $\lambda < L$. 
   
   The transition to localization with incresing the number of quantum wells 
    can be seen for the level spacing distribution of a fifteen 
   wells long superlattice, Fig. 5: (a) for the $n=12$ miniband and (b) for the 
   $n=11$ miniband. In this figure we also pay attention to the difference 
   in level spacing distribution for level at the center of a miniband, top of
   Fig.5, and at the edges, bottom of Fig.5.
   For $n=12$ states at the center of the miniband, top of Fig.5(a), the level 
   spacing distribution still follows a Wigner surmise. A clear transition to 
   a Poisson distribution is seen for the level spacing at the bottom edge of 
   the miniband (bottom of Fig.5(a)). The transition towards localization is 
   already seen for levels at the center of the $n=11$ related miniband.
   Due to 
   the inhomogeneous broadening of the quantized levels there is no miniband 
   formation in the usual sence of ordered systems which could be characterized 
   by a near clean or at least a balistic regime \cite{rauch,wacker}. 
   Nevertheless, minibands in a less strict sense are 
   defined by a set of states which repeal each other and are delocalized 
   through the whole system.
   
   These results suggest that 
   finite superlattices of quantum wells with correlated disorder present 
   very interesting properties related to the competition between the 
   level repulsion and the level broadening. 
   Such a competition can be defined for the fact 
   the inhomogeneous broadening is mainly related to the properties of the 
   repulsive binary alloy from which the quantum wells are made of, while the
   level repulsion is introduced by the coupling of the quantum wells,
    which can be partially tuned through 
   more or less transparent barriers. From this point of view, the transition 
   from diffusive-like regime to a localized one may be obtained by  
     means of three different mechanisms. 
     
     The first mechanism is the suppression of 
     level repulsion by increasing barrier widths, ilustrated for the 
     double-well case in Fig.3. By increasing the barrier width, the coupling
     between the wells is reduced and the states tend to localize in only one
      of the wells. Here we see that the tunneling range depends on the 
      inhomogeneous broadening of the almost resonant states. For barriers wider 
      than the tunneling range, the states become uncorrelated. In other words,
      varying the barrier width, introduces a spatial dependence of the density 
      correlation function in analogy to the energy-level correlation function 
      analysis for the Anderson disordered tight-binding model by Sivan and 
      Imry \cite{sivan}.

 A second mechanism of couplig suppression is simply related to the localization 
properties of the bulk chain that constitutes the 
material disordered quantum wells. Hence, for a given superlattice, there are 
minibands with different degrees of localization, from clusters of states with 
true minibands characteristics to completely localized ones, including at least 
one miniband with a diffusive character. This effect can be followed for the 
 different minibands of a five well superlattice in Fig.4. 
 
 Probably the most interesting repulsion suppression mechanism is related to 
 increasing superlattice length. 
  Related to this mechanism 
 is  the 
 expected behaviour 
 that a difusive-like miniband in a finite superlattice becomes 
 localized, if the number of quantum wells makes the superlattice exceed the 
 localiztion length for this given miniband. The level spacing statistics evolve
  from a Wigner surmise towards a Poisson distribution because some levels 
  start to become uncorrelated. This behaviour can be identified by 
  comparing Fig.4 with Fig.5. 
   
 On average, the level repulsion within a miniband decreases with the 
 increase of the number of quantum wells in a superlattice. 
 This follows the textbook picture \cite{bastard}, 
 within a tight-binding framework, 
 of the formation of continuous bands for infinite crystals by 
 the successive addition of atoms to the system.

\section{Motional narrowing}

Although the suppression of level repulsion with increasing
 the number of quantum wells can be identified in the level spacing statistics, 
 a more precise signature of this mechanism is related to a motional narrowing 
 effect. 
 The term "motional narrowing" has been used to 
 designate the reduction of spectral line widths in disordered systems by some 
 averaging process. Such effect occurs in many areas of physics and, in 
 particular, has recently been predicted and 
 observed in semiconductor microcavities \cite{whittaker}.
  In the present work, the motional narrowing shows up 
 as a diminution of the inhomogeneous broadening of the states with increasing 
 the system length. In Fig. 6 the average histograms of individual levels
 of the $n = 11$ miniband for superlattices of different length are shown;
  $N_w = 5$, Fig. 6(a); and $N_w = 9$, Fig. 6(b). It is noticeable the shrinking 
  of the average half width of the states with increasing the number of quantum 
  wells. For a better comparison, we pay special attention to the central 
  miniband state in either case. Actually the narrowing of the broadening is 
  more intense for this state. This narrowing may be qualitatively understood 
  by the fact that the level repulsion inhibits the inhomogeneous broadening of 
  an individual level taken as an average over many disorder configurations,
  since the levels in a finite superlattice are correlated. In other words, 
  coupling of states shrinks the range for broadening and the central state in 
  a miniband feels the strongest reduction of this broadening range. In a more 
  precise way, this motional narrowing effect is predicted by the RMT, having
  in mind the expression for the Wigner surmise. The average level repulsion,
  $D$, in eq. (3) is directly related to the half width of the random gaussian
  distribution of $H_{ij}$ in eq.(2), $D=\sigma\sqrt{{\pi\over 2}}$.  
  A finite disordered superlattice 
  is characterized, in  analogy to the 
  ordered counterpart, by the reduction of the average level repulsion 
  with increasing system size. This reduction leads to a diminution of the 
  broadening of the random distribution of the Hamiltonian matrix elements.
  Hence, the motional narrowing is an effect directly related to 
  and explained by 
  the RMT.

\section{Conclusions}

  The present model for a finite superlattice overcomes 
   the usual difficulties of RMT applied to one dimensional systems. 
   Furthermore, the transition from Wigner surmise to Poisson, 
   by increasing the
   system length, suggests that some minibands - although defined in a
   one-dimensional system - show properties of diffusive regimes. Important
   here is the presence of correlations, as well as the limit of states 
   completely isolated in a single quantum well and the natural localization 
   length selection by the miniband index. However, there is also a fine tuning
   of the localization length in each miniband, as can be seen already from the
   differences in level spacing distribution for different level pairs within a
   miniband, as depicted in Fig.5. Further, a less pronounced motional narrowing 
   for states at the edges of a miniband, compared to central ones, is also an 
   indication of this localization lentgh tuning. A definitive result for the 
   modulation of the localization length within a miniband is given by the 
   participation ratio \cite{thouless}

\begin{equation}
{\cal P}= \sum_{\vec{r}}{|\Psi(\vec{r})|^{4}}, 
\end{equation}

\noindent
where $\parallel \Psi \parallel=1$.
 In the tight-binding model with $N$ atomic sites eq.(5) becomes:

\begin{equation}
{\cal P}= \sum_{i=1}^{N}{|a_{i}^{\alpha}|^{4}} = {\cal A}_{\alpha}
\end{equation}

\noindent
where $a_{i}^{\alpha}$ is the $i$th site component of $\alpha$ eigenstate.
     
    The participation ratio, $P$, is inversely proportional to the localization 
   length \cite{izrailev96}.
   Besides, the participation ratio delivers an overview of the main
   results shown in the present paper.
   In Fig. 7, the average participation ratio for $N_w = 19$ 
   superlattices are shown. Open circles are for thin barriers, $L_b =2$ and 
   full squares for thick barriers, $L_b = 5$. For both cases, 
   several minibands separated by minigaps are identified. However, only for
   the thin barrier case, the minibands show a clear behaviour of the centre 
   less localized than the edges. 
   This recalls the mechanism of
   suppressing level repulsion by increasing 
   barrier width. We also observe the reduction of localization
   length together with increasing broadening for minibands progressively away 
   in energy from $n = 12$ miniband, recalling the behaviour of the localization 
   length of a repulsive binary alloy bulk chain. It should be noticed that for 
   energies below -0.85 eV and above -0.5 eV in Fig.7, no minibands are resolved 
   in the energy spectra. Nevertheless a clear modulation of the localization 
   length is still present. This suggests that minibands in disordered 
   superlattices could be still defined by a modulation of the localization 
   lentgh in a continuous spectrum.

\section{Acknowledgements}

R.R. Rey-Gonz\'alez 
gratefully acknowledges the hospitality of the Instituto de F\'isica 
Gleb Wataghin at the State University of Campinas, where part of this work was 
done. This stay was possible with the financial support from 
the Facultad de Ciencias of the Universidad 
Nacional de Colombia,
together with International Program in the Physical Sciences
 (Uppsala University)
 and the Funda\c{c}\~ao de Amparo \`a Pesquisa do Estado de
S\~ao Paulo (FAPESP). P.A.S. also acknowledges the financial support from FAPESP
 and the Conselho Nacional de Pesquisa e Desenvolvimento (CNPq). P.A.S. would 
 like to thank E. R. Mucciolo for useful discussions.

\begin{figure} \caption{Part of the energy spectra for 100 different disorder
configurations: (a), for a single quantum well; (b), for a double quantum well;
 and, (c), for a
finite superlattices of five quantum well. For other
structure parameters, see text.}  \label{1} \end {figure}

\begin{figure} \caption{(a) Histograms of energy distributions for the lower
 and upper level of the n=12 doublet of a double disorder
quantum well.  (b) Average density of states for the
same n=12 doublet in (a). Quantm well parameters are the 
same as in Fig. 1(b).}  \label{2} \end{figure}

\begin{figure} \caption{Nearest level spacing distribution for the n=12,
(a), and n=11, (b) doublet of a double disorder quantum well for different
central barrier thickness:  Top for thin, $L_b=2$ sites, barriers; and bottom 
for thick,
$L_b=5$ sites, barriers.  Poisson distribution (dashed lines) and 
Wigner (dotted lines) surmise
are also shown for comparision.}  \label{3} \end{figure}

\begin{figure} \caption{Nearest level spacing distribution for n=12,
(a), n=11, (b), n=10, (c), and n=9, (d), minibands of a finite superlattices of
five quantum wells.  Poisson distributions(dashed lines) and Wigner 
(dotted lines) surmises, for the respective average nearest level spacings, 
are also shown for comparision.}  \label{4} \end{figure}

\begin{figure} \caption{Nearest-neighbour level spacing distributions
 for the n=12,
(a), and n=11, (b), minibands of a fifteen quantum
wells finite superlattice. Top: distribution for central levels in the miniband. 
Bottom: distribution for lewels at the bottom of the miniband. 
 Poisson distributions (dashed lines) and Wigner (dotted lines)
surmises are also shown.}  \label{5} \end{figure}

\begin{figure} \caption{Energy histogram for individual
levels of the n=11 miniband for finite superlattices: (a) five quantum wells;
 and (b), nine quantum wells. Thick lines corresponds to the center
level in these minibands.}  \label{6} \end{figure}

\begin{figure} \caption{Participation Ration as a function of energy for a
finite superlattice of nineteen disordered quantum wells. 
Different barrier thicknesses are considered: $L_b=2$ sites (open circles), and 
$L_b=5$ sites ( full squares).}  \label{7} \end{figure}

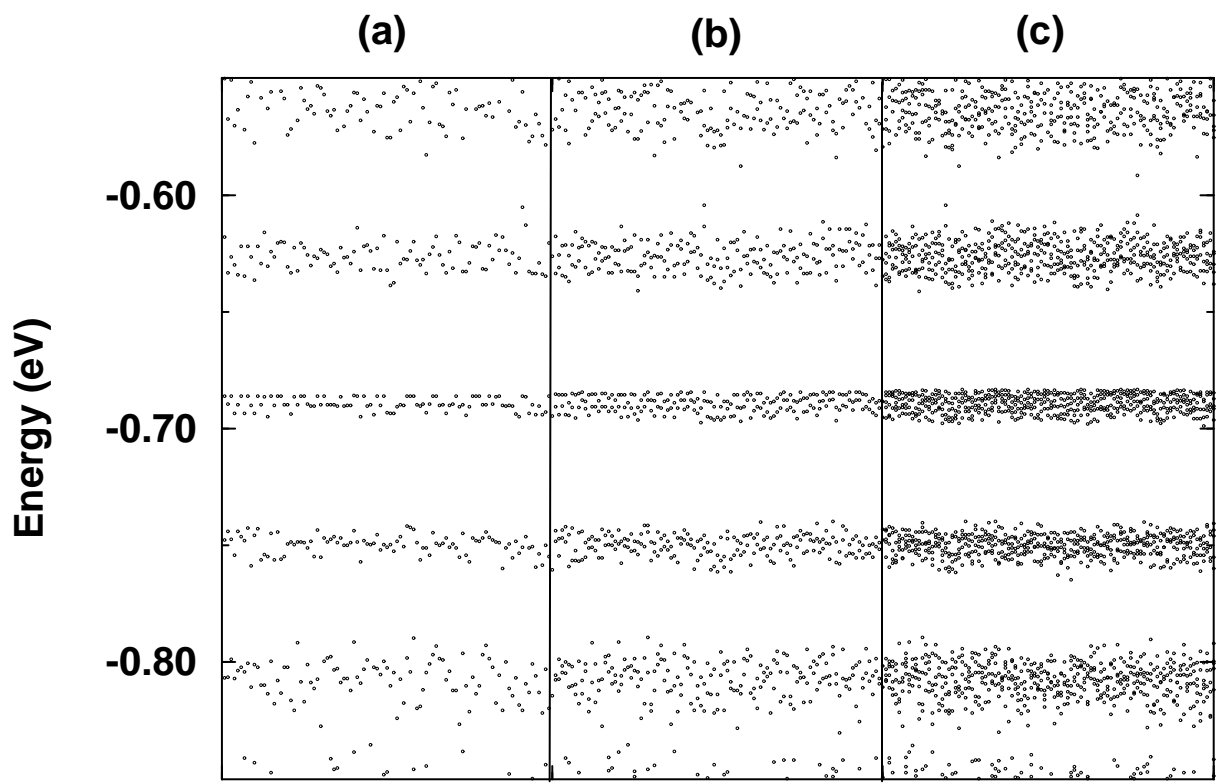
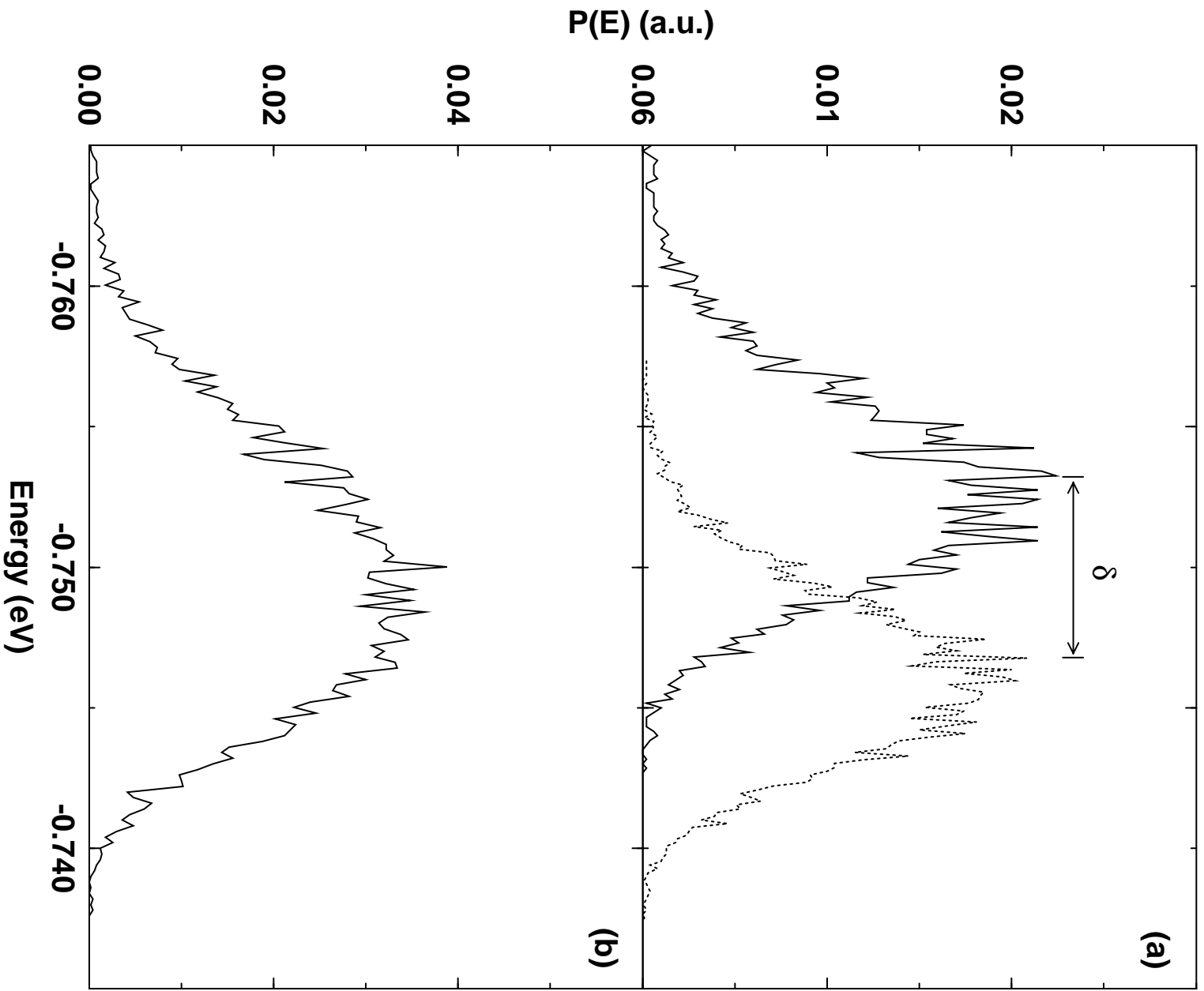
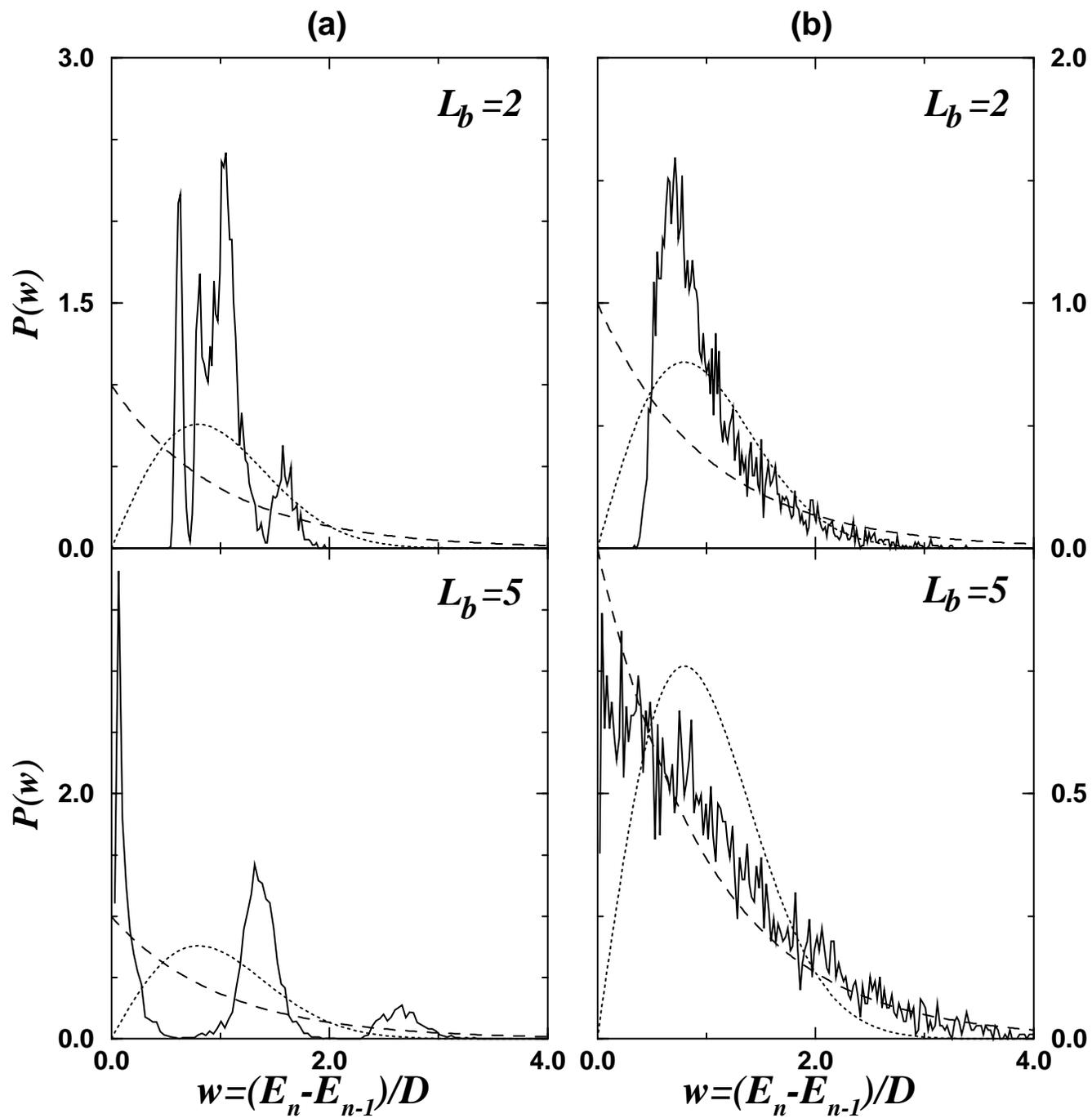
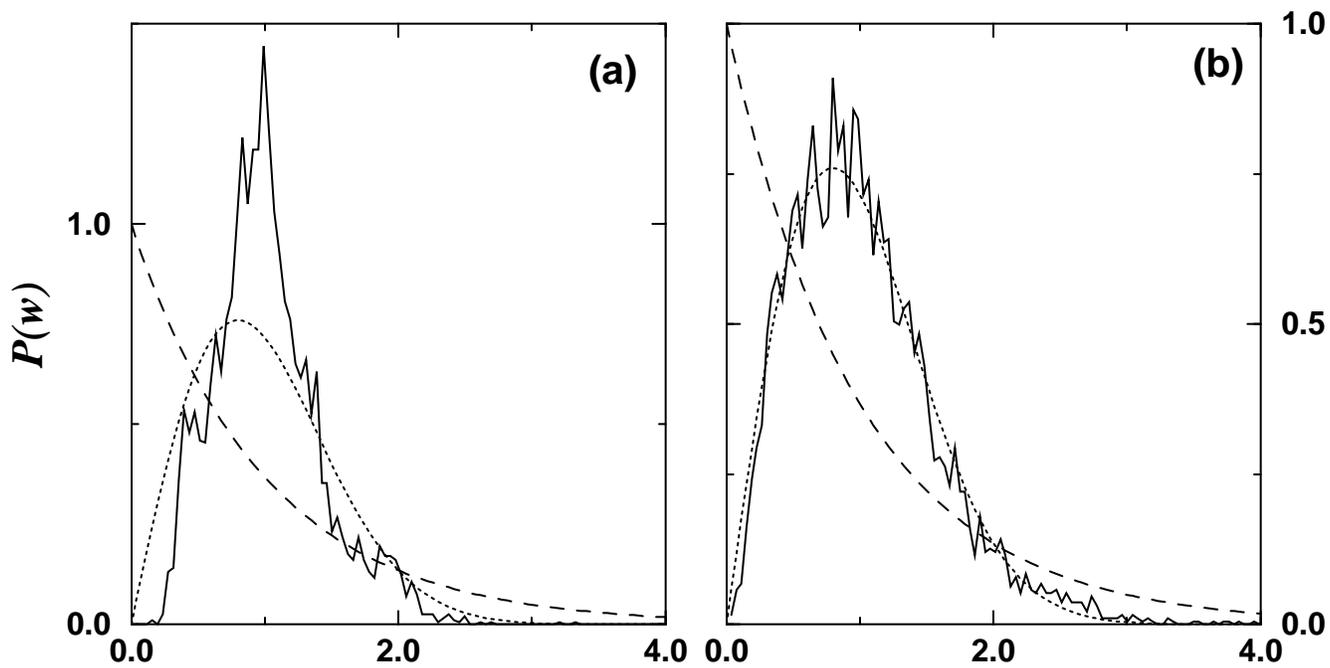
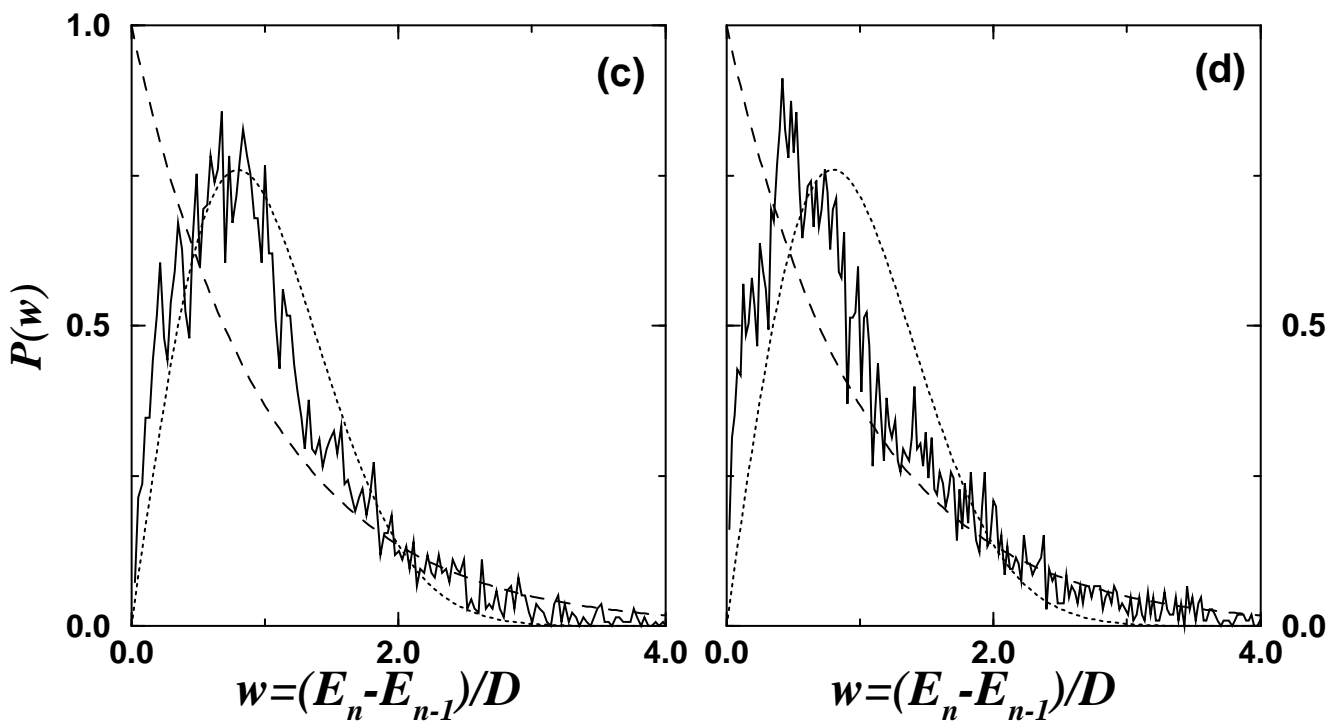
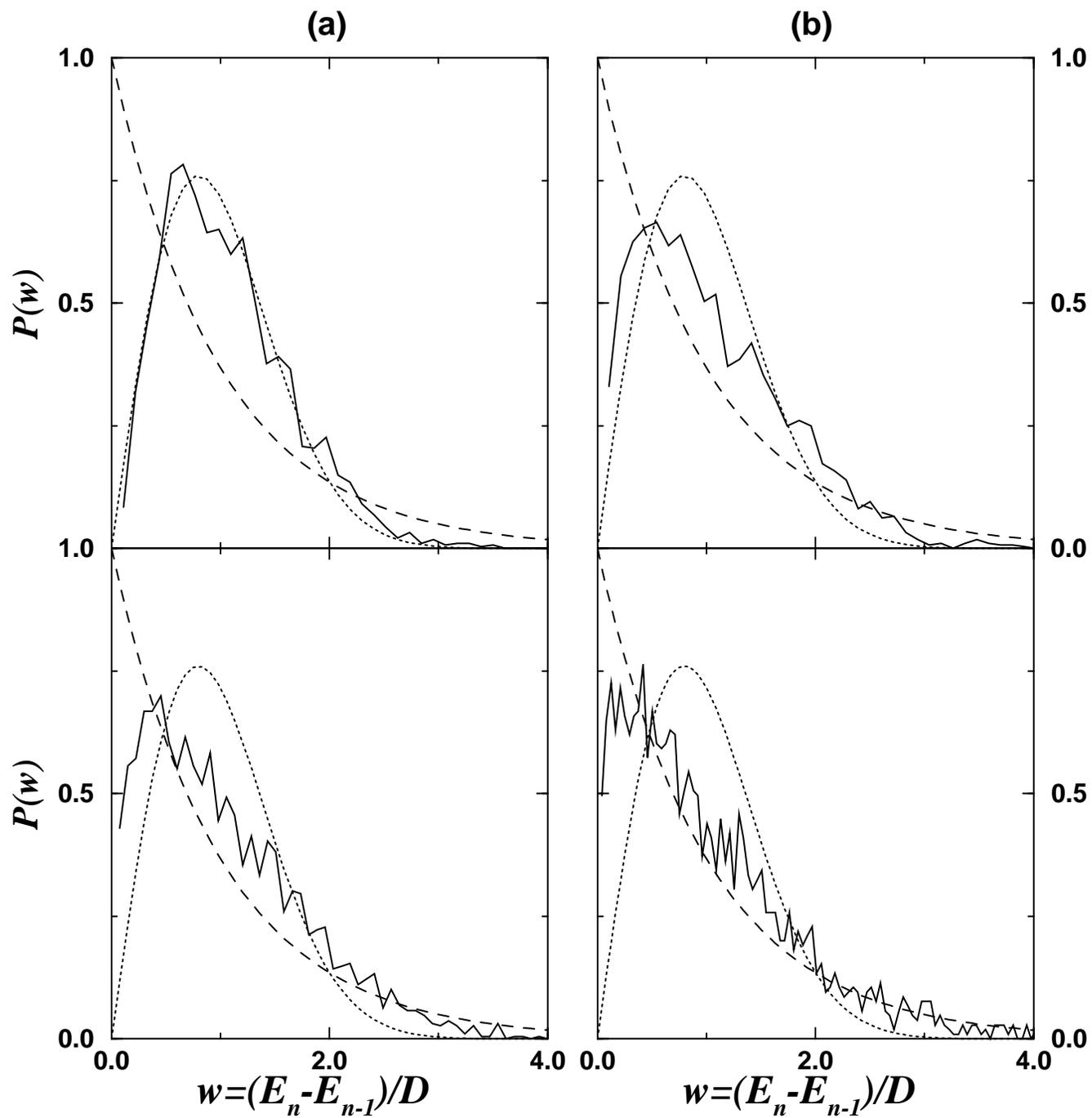
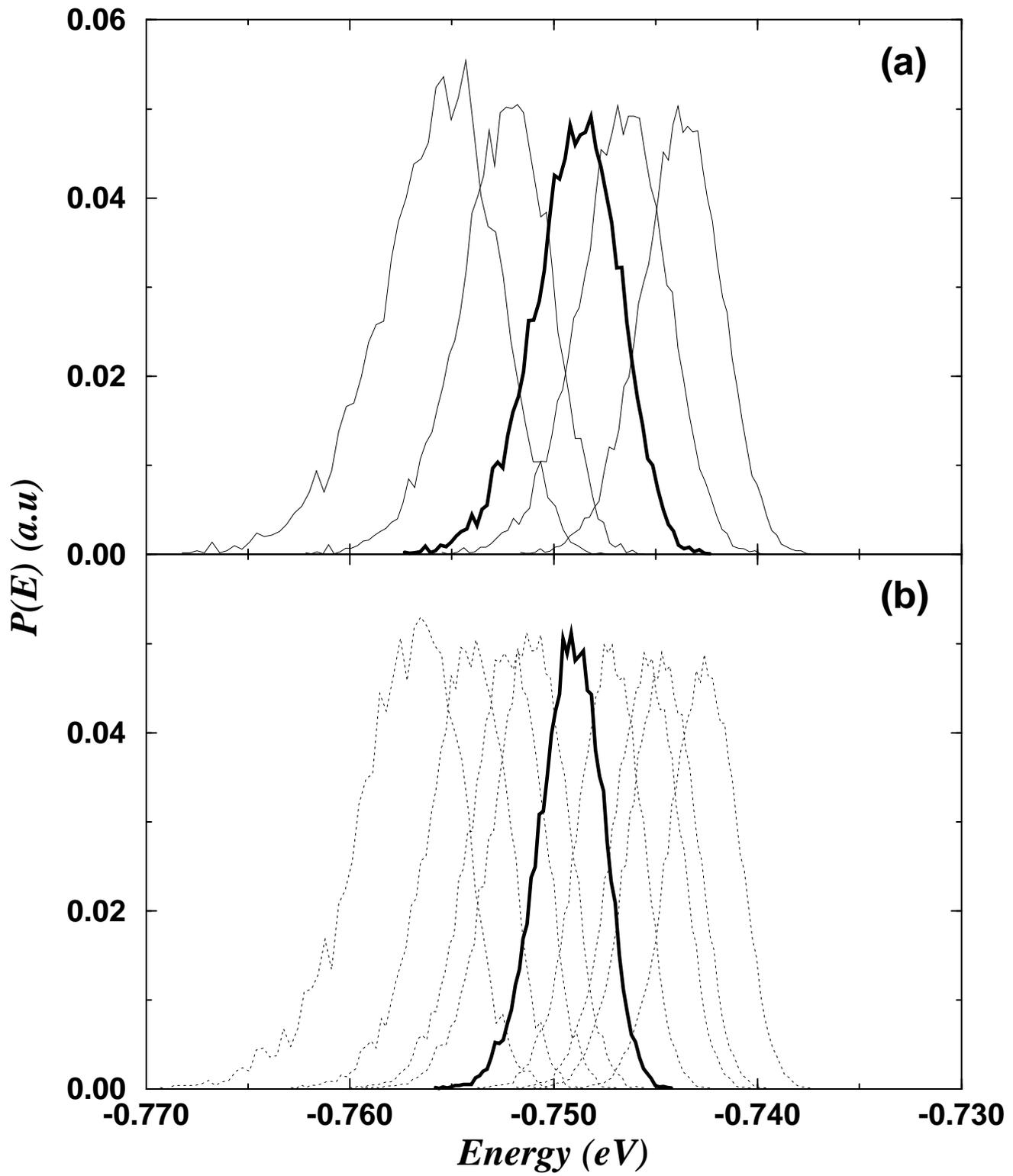
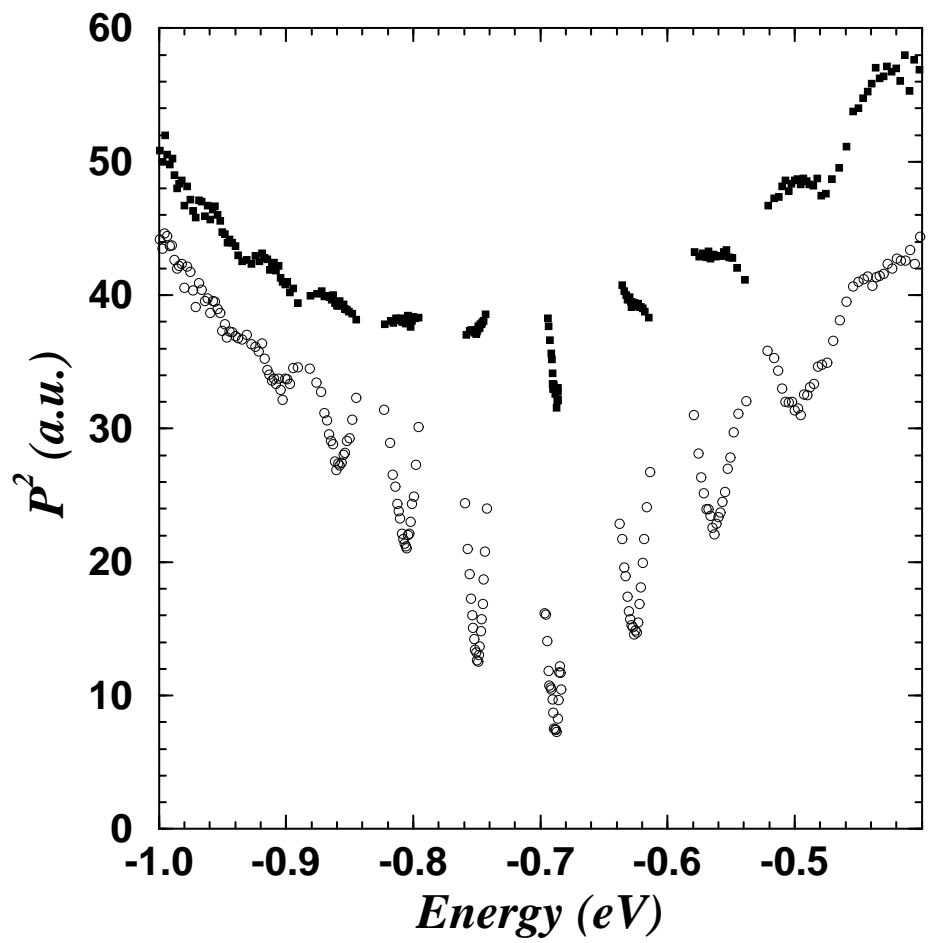
\end{document}